\documentclass[conference]{IEEEtran}
\pdfoutput=1
\usepackage[nolist]{acronym}
\usepackage{amsmath, bm}
\usepackage{amssymb}
\usepackage[table,xcdraw,dvipsnames]{xcolor}
\usepackage{algpseudocode}
\usepackage{balance}
\usepackage{tabularx}
\usepackage{subcaption}
\usepackage{multirow,array}
\usepackage{url}
\usepackage{textcomp}
\usepackage[linesnumbered,ruled,vlined]{algorithm2e}
\usepackage{amsmath}
\usepackage{amsfonts}

\usepackage{lipsum}
\usepackage{stfloats}
\usepackage{easyReview}
\ifCLASSINFOpdf
\usepackage[margin=0.9in]{geometry}
\fi
\begin{document}
\title{Dynamic Prioritization and Adaptive Scheduling using Deep Deterministic Policy Gradient for Deploying Microservice-based VNFs\\}

\author{\IEEEauthorblockN{Swarna B. Chetty, Hamed Ahmadi
\IEEEmembership{Senior Member,~IEEE}, Avishek Nag \IEEEmembership{Senior Member,~IEEE}}

}

\maketitle
\bstctlcite{IEEEexample:BSTcontrol}
\begin{abstract}
The \ac{NFV}-\ac{RA} problem is NP-Hard. Traditional deployment methods revealed the existence of a starvation problem, which the researchers failed to recognize. Basically, starvation here, means the longer waiting times and eventual rejection of low-priority services due to a `time out'.
The contribution of this work is threefold: a) explain the existence of the starvation problem in the existing methods and their drawbacks, b) introduce \ac{AdSch} which is an `intelligent scheduling' scheme using a three-factor approach (priority, threshold waiting time, and reliability), which proves to be more reasonable than traditional methods solely based on priority, and c) a \ac{DyPr}, allocation method is also proposed for unseen services and the importance of macro- and micro-level priority. 
We presented a zero-touch solution using \ac{DDPG} for adaptive scheduling and an online-\ac{RR} model for dynamic prioritization. The \ac{DDPG} successfully identified the `Beneficial and Starving' services, efficiently deploying twice as many low-priority services as others, reducing the starvation problem. Our online-\ac{RR} model learns the pattern in less than 100 transitions, and the prediction model has an accuracy rate of more than 80\%.

\end{abstract}
\section{Introduction}
Although \ac{5G} is presently providing the fundamental support for the \ac{IoE} and \ac{URLLC}, it is debatable if the current \ac{5G} systems can smoothly handle applications like \ac{DT}, connected robotics, autonomous systems, \ac{AR}/ \ac{VR}/ \ac{MR}, and Blockchain and Trust technologies ~\cite{saad2019vision, DTpaper21}. These upcoming applications are envisioned to request services with stringent standards, such as high reliability, low latency, and significant data rates~\cite{saad2019vision}. Due to this debate, there has been a significant research progress towards \ac{6G}.
The \ac{6G} must be tailored to support the upcoming service types like \ac{COC}, \ac{CAeC}, and \ac{EDuRLLC} in addition to \ac{eMBB}, \ac{URLLC}, and \ac{mMTC} services~\cite{letaief2019roadmap}.

There are several initiatives on both the access and the network side to facilitate the transition towards the \ac{6G}.  On the network side, the \ac{NFV} architecture, introduced in 2012~\cite{NFVproposal}, is pushing towards the microservices-based architecture~\cite{9306098,chowdhury2019re}. 
The \ac{NFV} framework virtualizes the \acp{NF} from their dedicated proprietary substrate appliances by allowing them to run as softwarized \acp{NF} (say, \acp{VNF}) on commodity hardware. This enables freedom, flexibility, and agility for \acp{VNF} to migrate from one server to another in response to dynamic variations in resource demand. Although \ac{NFV} is a promising technology, it can be challenging when various applications coexist, and simultaneously the underlying infrastructure requires guaranteed resources for all the arriving \ac{SFC}\footnote{Usually, multiple \acp{VNF} that are required by a service are `chained' in the order in which they are accessed by a service. This is called a \ac{SFC}.}. This is an NP-Hard problem and is known as the \ac{NFV}-\ac{RA}. 
In this type of \ac{NFV}-\ac{RA}, due to the affinity and anti-affinity constraints\footnote{Affinity and anti-affinity constraints refer to the ability to embed a \ac{VNF} on a particular node (affinity) and the converse of it (anti-affinity)}, the complexity grows further, hindering effortless software updates and routine maintenance. To address this, microservices offer an increased degree of freedom in the scalability, flexible upgrades and maintenance of \acp{VNF}. 
This cutting-edge `\ac{NFV}-\ac{RA} + Microservies' strategy ppromises to offer a solution  (theoretically) closer to the optimum, despite having an intensified design and deployment complexity. 
In \cite{9923918} we provided a detailed analysis of this approach and the criteria for executing dynamic decomposition, of which, Section~\ref{subsec:4} provides an overview.

We examine a deeper analysis of the criteria for dynamically prioritizing the online \acp{SFC}, the merits of embedding an \ac{SFC} over others and the significance of admission control. Most research to date has concentrated exclusively on deploying \acp{SFC} using \ac{FIFO}. Realistically, not all \acp{SFC} fall under the same priority group and cannot be addressed equally. The traditional method, \ac{FIFO}, failed to recognize and give emergency \acp{SFC} higher preference than non-emergency ones, causing the rejection of highly critical \acp{SFC}. In a modified approach, the priority-based approach is established to prefer the high-priority \acp{SFC} first over the lower ones~\cite{cappanera2019vnf, mohamad2020psvshare}, which overcomes the \ac{FIFO} drawback. Nevertheless, this introduced `starvation for lower-priority services, causing a significant biasness in the system - an unfair design to the users. To dismiss this biasness, the critical \ac{SFC} should not be classified based on priority levels but rather with other essential attributes. Thus, requiring an `intelligent fair system'. Moreover, with the lack of requested service information, predicting the types of arriving \acp{SFC} and their priority level will be challenging considering future communications. Thus, we need to dynamically classify the online services. 

In this paper, to provide seamless support to more realistic services for future networks, we propose a \ac{DyPr} and an \ac{AdSch} module for arriving \acp{SFC} using the \ac{RR} model and \ac{DDPG} approach, respectively, to diminish the `Starvation’ situation and provide a fair scheduling principle. Our proposed \ac{DDPG}-based algorithm identifies and understands the importance of embedding the `Beneficial SFC’ or `Starving SFC’ over others. The criteria and methods for assessing the priority of the online service have also not been specified in the literature; instead, a static approach has been adopted. Thus we have first proposed a standard for estimating the priority of an \ac{SFC} dynamically by using the \ac{ML} technique. Using the achieved priority and other \ac{QoS} attributes, we trained our \ac{DDPG}-based model to rank the arriving \acp{SFC} based on their urgency and requirements. Based on the \ac{SFC}’s rank, the rescheduling for deployment occurs; later on, the \ac{SFC} has to go through the admission control module, which provides permission or rejection for deployment based on the waited time by the \acp{SFC} in the queue. 
In summary, the main contributions of this paper are:
\begin{itemize}
    \item Establish a dynamic priority estimation criteria for all the online services using the \ac{RR} model.
    \item Develop a \ac{DDPG}-based framework for recognizing and understanding the importance of preferring an \ac{SFC} (say, `Beneficial and Starving \ac{SFC}’) over others.
    \item Establish an admission control approach, based on the \ac{SFC}'s waiting time in the queue before being deployed.
    \item  Adapting the deep \ac{QL} model along with microserives concept for \ac{VNF} embedding. 
\end{itemize}
\section{Literature review}\label{sec:2}
To meet the \acp{SLA} of next generation networks, \ac{NFV} has received a lot of attention. The majority of the research has focused on the placement of arriving \acp{SFC} in a \ac{FIFO} fashion. For instance, \cite{mijumbi2015design, agarwal2019vnf} formulated the problem based on the exact optimization method; however, the solutions are delivered using heuristic or meta-heuristic models due to the high computational cost. \ac{ML} techniques are commonly proposed as an effective technique for handling complicated problems like \ac{NFV}-\ac{RA}. Authors in~\cite{yuan2020q, sciancalepore2018z, quang2019deep, chetty2020virtual, chetty2021virtual} proposed reinforcement learning-based approaches. In a realistic scenario, each service has its own level of importance based on its type and requirements. Following the \ac{FIFO} technique can cause failure in deploying emergency services over best-effort ones. 
Methods that classify \acp{SFC} as either \ac{Pr} or \ac{BE}, constantly prioritizes \ac{Pr} services over the \ac{BE} ones, resulting in severe service deprivation for low-priority services~\cite{cappanera2019vnf}. \cite{malandrino2019reducing} and \cite{jalalitabar2016service} demonstrated the effectiveness of mapping the services depending on the priority of \acp{VNF}. The authors of \cite{malandrino2019reducing} discuss three types of priority: per-service, per-\ac{VNF}, and per-flow, where service prioritization is based on the delay attribute and per \ac{VNF} priority is assigned randomly. This is an iterative process that adjusts the solution after each iteration, which is not scalable considering future requirements. Also the predefined number of flows that a \ac{VNF} can handle, ignores the operational dynamics.

Priority-based scheduling and deployment is still an issue requiring an `intelligent' system to dynamically establish the online service priority and re-schedule the services according to the urgency and benefit. Researchers have proposed \ac{ILP} or heuristic models in the literature, which are either computationally costly or produce sub-optimal solutions. To overcome this drawback, we propose \ac{ML}-based algorithms to dynamically assign priority to online services and train the model to re-arrange the scheduling queue according to the needs. 

\section{System Model}\label{sec:3} 
The physical topology is modelled as a directed graph $\textbf{G} = (\textbf{H}, \textbf{N})$, where $\textbf{H} = [\textbf{h}_{0},\ldots,\textbf{h}_{|H|-1}]$ and $\textbf{N} = [\textbf{n}_{0},\ldots,\textbf{n}_{|N|-1}]$ represent the set of physical nodes (say high-volume servers) and physical links of the topology, respectively.  Each physical node $\textbf{h}_{x'} = [h_{x', 0},\ldots, h_{x', J_{node}-1} ]$ denotes the amount of available nodal resources like CPU core, RAM, etc, and $J_{node}$ is the number of nodal resource types indexed from $0, 1,\ldots,J_{node}-1$. Similarly, each physical link $\textbf{n}_{y'} = [n_{y',0},\ldots,n_{y',J_{link}-1}]$ signifies the amount of link resources like bandwidth, latency, etc. $J_{link}$ stands for the number of link resource types, the available resource for physical link.
The \ac{VNF-FG} or \ac{SFC}~(${\Psi}$) is represented as a directed graph $\textbf{G}_{\Psi}'=(\textbf{V}_{\Psi},\textbf{B}_{\Psi})$ with the set of \acp{VNF} ($\textbf{V}_{\Psi} = [\textbf{v}_{\Psi, 0},\ldots,\textbf{v}_{\Psi, |V|-1}]$) and \acp{VL} ($\textbf{B}_{\Psi} = [\textbf{b}_{\Psi, 0},\ldots,\textbf{b}_{\Psi, |b|-1}]$) , that delivers an end-to-end service. Deploying these graphs onto the physical topology is termed as the \ac{VNF-FGE} problem. Each \ac{VNF} ($\textbf{v}_{\Psi, x} = [r_{\Psi,x,0},\ldots,r_{\Psi,x,J_{node}-1}]$) and \ac{VL} ($\textbf{b}_{\Psi, y} = [l_{\Psi,y,0},\ldots,l_{\Psi,y,J_{link}-1}]$) comprises of set of requested resources like CPU core, delay, bandwidth etc. 
The computing resources initialization in most related works, is done in a random manner, disregarding the high correlation between the CPU core and RAM. This work outlines the link between the CPU core and RAM as in~\cite{gupta2018service}. In addition, we also considered delay, jitters, packet loss, reliability and threshold waiting time as some of the \ac{QoS} attributes, thus making the $J_{node} = 6$ in our case. Similarly, $J_{link} = 2$ in our studies, considering latency and bandwidth. 


\subsection{Dynamic Prioritization}\label{subsec:1} 
A major drawback of commonly-used methods of binary classification of services into \ac{Pr} and \ac{BE} is that, the pre-determined priority levels for the arriving services are randomly allocated and the co-relation between the service requirements and priority are ignored. In a realistic scenario, the allocation of service priority should be based on various factors of QoS, flow type, etc. Determining the priority levels for the existing or well-aware services is trivial. The complexity induces when future communications system expects frequent unseen `short-lived' services with instant placement requirements. To fulfil this need, an intelligent and \ac{DyPr} model is anticipated rather than a static or conventional method.
In the \ac{DyPr} model, we investigate the relationship between the requested \ac{QoS} factors of the arriving service and, based on it, a dynamic priority level is assigned. This problem is viewed as a multiple regression task with various independent variables $A[0]\dots A[p]$ (like \ac{QoS} factors with $p$ as the number of factors), to predict a dependent continuous variable $Y'$ as referred in Eqn. (\ref{eq:multiR}). Here $W$ and $B$ are the regression coefficient and residual term respectively, which are learned during the training. The adopted model discovers the linearity between the dependent and independent variables. In our case, these independent variables are threshold jitters, delay, and packet loss, which a service can tolerate, and the dependent variable denotes the appropriate priority level. However, the multiple regression data suffers from multi-colinearity, which causes the estimation of regression coefficients to be inaccurate. As a result, its existence reduces the model’s performance by causing the predicted value to differ significantly from the actual value. One way to diminish this is by adopting the \ac{RR} model, which performs L2 regularization, due to which the model is more restricted and causes less over-fitting \cite{introML}. The objective is to minimize the cost function mentioned in~(\ref{eq:costRid}), where $M$ is the number of samples (services per run), $Y_\Psi$ is the actual value.  The $\lambda$ (penalty term) regularizes the coefficients in such a way that the optimization function is penalized if the coefficients assume high values. 
\begin{equation}\label{eq:multiR}
   \footnotesize Y'_\Psi = W[0]\times A_\Psi[0]+ \dots + W[p]\times A_\Psi[p] +B
\end{equation}
\vspace{-0.189in}
\begin{equation}\label{eq:costRid}
   \footnotesize \sum_{\Psi = 0}^{M} (Y_\Psi - Y'_\Psi)^2 = \sum_{\Psi = 1}^{M} (Y_\Psi - \sum_{j = 0}^{p}W_j\times A_{\Psi,j})^2 + \lambda \sum_{j = 0}^{p} W_{j}^2
\end{equation} 

\begin{figure}[h]
    \centering
    \includegraphics[width=0.75\textwidth]{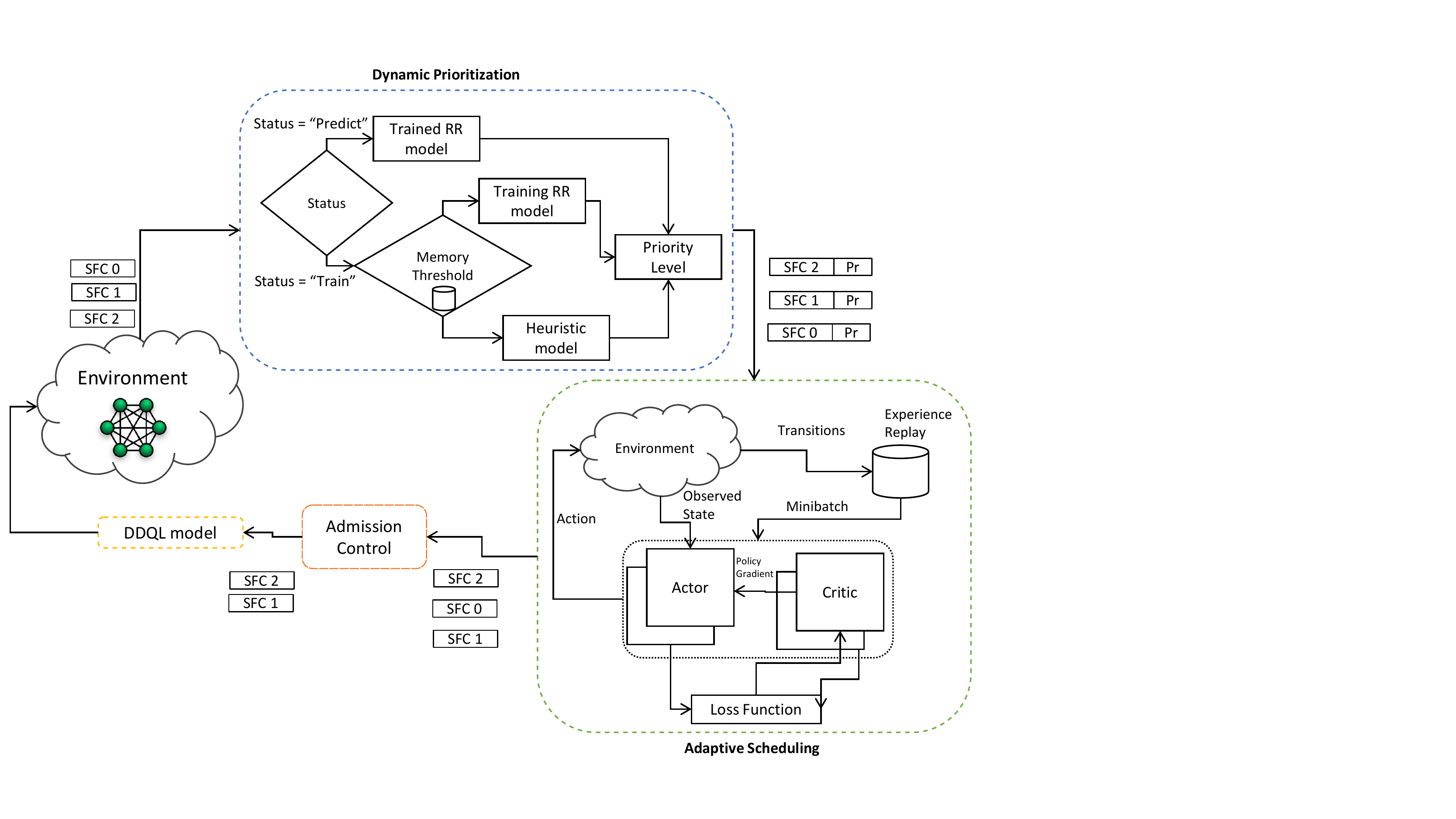}
    \caption{Overview of the \ac{DyPr} scheme.}
    \label{fig:Overview}
\end{figure}

The \ac{RR} model is supervised learning, which uses pre-existing datasets for training purposes. Our study evaluates the model's performance for future unseen services; hence, we assume no helpful service information is available to us. The traditional \ac{RR} model would not provide enough support due to the lack of sufficient datasets. To support the dynamism, we have modified it by adding an observation phase before the learning phase. This helps in constructing our training datasets, making an `Online-\ac{RR}' model. In the observation phase, due to the unavailability of service information, the model saves (observes) the online services in a memory buffer until the minimal transitions is achieved to begin the training phase. During this time, the allocation of priority is performed uniformly. Later, once the saved transition threshold is surpassed, the learning phase starts, which uses the current and saved transitions for model training. These saved transitions (batch transitions) are selected randomly from the memory to avoid any chance of overfitting (co-relation between the transitions). The model's accuracy is checked periodically. Once the model reaches the desired accuracy, the trained model state changes from `Train' to `Predict'. The overview of \ac{DyPr} is shown in Fig.~\ref{fig:Overview}.
Moreover,  this model was chosen after a thorough evaluation of model performances using several methodologies, including \ac{ANN} and Lasso. Of all the models, the online-\ac{RR} model outperformed expectations. This model is constructed with an envision to support future unseen services by introducing a zero-touch cognitive system. 
Traditionally, the services are categorized into two cardinalities, which diminishes the value of services relative to one another within a class. Most research has ignored the intra-class priority aspect; however, given the future demands, it should be considered. To overcome this, a service priority is ranked between 0.0 to 1.0, with
0.0 represents the least essential service, and 1.0 being the most crucial. Each priority level is expressed in macro-class priority and micro-class priority. While the micro-class priority establishes the priority status inside a class, the macro-class priority categorizes the class to which it belongs. For example, the priority of \ac{SFC} A and \ac{SFC} B is 0.79 and 0.72, respectively. Though both the \acp{SFC} belong to the same class (i.e., macro-class is 7), \ac{SFC} A with a micro-class as 9 will be prioritized over the \ac{SFC} B with a micro-class as 2, due to the higher value. 
This delivers more details about the importance of services within the class, which is beneficial, especially for time-Sensitive or critical applications and for the adaptive scheduling phase. Therefore, in our work, the \ac{DyPr} is considered as a Regression model rather than classification. 


\subsection{Adaptive Scheduling}\label{subsec:2} 
Conventionally, high-priority services are preferred over others, leaving the low-priority services to wait for extended periods of time, resulting in their starvation. This highlights the second drawback of conventional binary classification of services. A biased scheduling system, which raises the questions like, `Can the priority attribute be the most effective scheduling decision-making factor?' `Will the decision be more optimal by considering additional factors, such as service waiting time or reliability?' 
%
In a search for an unbiased scheduling system, we have proposed an intelligent Adaptive Scheduling module using the \ac{DDPG} approach. This approach weighs more than one factor and provides an optimal decision. Let us say, that a high-priority \ac{SFC}~A has a higher waiting threshold than a lower-priority \ac{SFC}~B (which is about to expire soon). According to the traditional model, \ac{SFC}~A will be selected, ignoring the \ac{SFC}~B to expire, affecting the \ac{QoE}. However, our model, which considers more than one factor, prefers \ac{SFC}~B over \ac{SFC}~A, understanding that there will not be any negative impact on \ac{SFC}~A if it waits in a queue a little longer, providing scheduling optimality.

\ac{DDPG} is an actor-critic \ac{RL} model, trained  
to identify `Beneficial and Starving' services based on 3-factors: \ac{TWT}~$(T)$, Reliability~$(\Gamma)$, and Priority~$(P)$ of the service.   
The state-space for the requested service $\Psi$ is represented as $(T_{\Psi}, \Gamma_{\Psi}, P_{\Psi})$.
Based on these factors, the \ac{DDPG} agent determines a rank for each service (i.e., action-space~$(A_{\Psi})$), indicating the significance of deploying the service. With a higher rank, the necessity to deploy the service is significant, resulting in a rank-based scheduling method. 
%
In order to train the \ac{DDPG} model, an appropriate reward function is essential since it provides feedback to the agent based on the given action's effectiveness. 
Equation (\ref{eq:rDDPG}) describes the reward function $R(\cdot)$, which comprises two parts: Beneficial-cost $(\Upsilon)$ and Starvation-cost $(\Phi)$, providing a trade-off between the high-priority, starvation, and high-reliable services. The $R(\cdot)$ is a point-based function; depending upon the significance of factors, the reward points $(\theta_{pts})$ are scaled, as in Eqns. (\ref{eq:bencost}) and (\ref{eq:starcost}), where $n(\cdot)$ is the normalized function.
\begin{equation}\label{eq:rDDPG}
   \footnotesize R(\Psi) = \Upsilon_\Psi + \Phi_\Psi,
\end{equation}
\begin{equation}\label{eq:bencost}
    \footnotesize \Upsilon_\Psi = [(1 -n(T_{\Psi})\times\theta_{pts}) + (\Gamma_{\Psi}\times\theta_{pts}) + (P_{\Psi}\times\theta_{pts})],
\end{equation}
\begin{equation}\label{eq:starcost}
   \footnotesize \Phi_\Psi = \zeta_{\Psi}\times\theta_{pts},
\end{equation}
\begin{equation}\label{eq:starRate}
    \footnotesize \zeta_{\Psi} = 
    \begin{cases}
      \alpha(1-\epsilon)^\kappa,& \text{if } Z_{\Psi} = 1\\
       0,              & \text{otherwise},
    \end{cases}
\end{equation}
\begin{equation}\label{eq:potentialS}
     \footnotesize  Z_{\Psi} = 
    \begin{cases}
      1,& \text{if } P_{\Psi} \leq 0.2\\
       0,              & \text{otherwise},
    \end{cases}
\end{equation}
Equation (\ref{eq:starcost}) introduces the biasness to diminish the starvation issue, where $\zeta_{\Psi}$ represents the starvation factor, which decays exponentially with the \ac{SFC}'s rank. 
According to Eqn. (\ref{eq:potentialS}), our algorithm determines if the arriving service qualifies as a `Potential Starving' service or not. On affirmative, the algorithm checks its placement position $\kappa$ in the adaptive scheduling queue, which impacts the starvation cost. $\kappa$ is determined using the exponential decay formula, with $\alpha$ as 1 and decay rate ($1-\epsilon$) as 0.1. The decay rate induces greediness in the system when the service is positioned at the beginning by giving higher rewards than others. Thus, making the agent position the starving services at the beginning. 

\subsection{Admission Control}\label{subsec:3} 
After achieving an optimal scheduling queue, we constructed an admission-control model to evaluate the extent to which the `Potential Starving' and `High-priority' services are deployed before expiration. This determines the trade-off between starvation and traditionally-preferred services. 
When a service is under placement, the waiting period $(\Delta)$ for the reminder services is recorded, which is illustrated in a 2-D matrix as below. A scheduling queue, for instance, is $[SFC_0, SFC_1, SFC_2, SFC_3]$.  When $SFC_0$ is placed, the remainder services' waiting span is $x_0$, which is the deployment time of $SFC_0$. With the deployment of $SFC_1$, the waiting time for $SFC_2$ and $SFC_3$ is increased by $x_1$ and so on. $\Delta_{\Psi,\Psi}$ depicts the total waited time by the service $\Psi$ in the queue. 
 \[\scriptsize
   \Delta = 
     \bordermatrix{ & SFC_0 & SFC_1 & SFC_2 & SFC_3  \cr
       SFC_0 & \Delta_{0,0} & \Delta_{0,1} & \Delta_{0,2} & \Delta_{0,3} \cr
       SFC_1 & \Delta_{1,0} & \Delta_{1,1} & \Delta_{1,2} & \Delta_{1,3} \cr
       SFC_2 & \Delta_{2,0} & \Delta_{2,1} & \Delta_{2,2} & \Delta_{2,3} \cr
       SFC_3 & \Delta_{3,0} & \Delta_{3,1} & \Delta_{3,2} & \Delta_{3,3}} \qquad
 \]
 \[\scriptsize
   \Delta = 
     \bordermatrix{ & SFC_0 & SFC_1 & SFC_2 & SFC_3  \cr
       SFC_0 & 0 & x_{0} & x_{0}       & x_{0}       \cr
       SFC_1 & 0 & x_{0} & x_{0}+x_{1} & x_{0}+x_{1} \cr
       SFC_2 & 0 & x_{0} & x_{0}+x_{1} & x_{0}+x_{1}+x_{2} \cr
       SFC_3 & 0 & x_{0} & x_{0}+x_{1} & x_{0}+x_{1}+x_{2}} \qquad
 \]
To initiate the placement, the service must satisfy the requirement as in Eqn. (\ref{eq:AC}).

\vspace{1.5in}
\begin{equation}\label{eq:AC}
       \footnotesize \Xi_{\Psi} = 
    \begin{cases}
      1,& \text{if } \Delta_{\Psi,\Psi} \leq T_{\Psi}\\
       0,              & \text{otherwise},
    \end{cases}
\end{equation}

\subsection{VNF-FGE Problem Formulation}\label{subsec:4} 
Result of \cite{9923918} show that the \ac{DDQL} model solves the \ac{NFV}-\ac{RA} problem efficiently, intending to embed maximum \acp{SFC} onto the substrate network under certain defined constraints. 
In \cite{9923918} we considered the physical topology as an environment comprised of high-volume servers. Each \ac{VNF}-\ac{FG} and its resource requirements are represented as state-space ($S$). The amount of physical nodes/servers present in the topology is described as the action-space ($A$). The Local reward function (i.e., the Eqn. (13) in~\cite{9923918}) has been modified as Eqn. (\ref{ConlocR}) , which is constructed based on the four attributes: 

\begin{enumerate}
    \item $R_{quality,y}$, the quality of the selected node ($y$) for deploying the \ac{VNF}, ($x$) is the ratio of available resources ($Ar_{y}$) in the node to the initialized resources ($Ir_{y}$), as in Eqn. (\ref{eq:Qnode}). The availability of resources in a physical node determines its quality.
        \begin{equation}\label{ConlocR}
            \footnotesize   L_{reward}(x)= 
            \begin{cases}
               \boldsymbol{R_{vnf}},& \text{if } \Phi_{x}^{y} = 1\\
               P^{pt}_{vnf},              & \text{otherwise}
            \end{cases}
        \end{equation}
        \begin{equation}\label{eq:Lreward}
           \footnotesize \boldsymbol{R_{vnf}} = R_{quality,y} + R_{priority} + R_{rel} + R_{placement}
        \end{equation}
        \begin{equation}\label{eq:Qnode}
           \footnotesize R_{quality, y} = \frac{Ar_{y}}{Ir_{y}}\times R^{pt}_{vnf}
        \end{equation}
        \begin{equation}\label{eq:RPr}
           \footnotesize R_{priority} = P_{\Psi}\times R^{pt}_{vnf} 
        \end{equation}
        \begin{equation}\label{eq:Rrel}
           \footnotesize R_{rel} = \Gamma_{\Psi}\times R^{pt}_{vnf} 
        \end{equation}
        \begin{equation}\label{eq:Rplacement}
          \footnotesize  R_{placement} = J^{x}_{y}\times R^{pt}_{vnf} 
        \end{equation}
        
    \item Based on the service priority ($R_{priority}$), as in Eqn. \eqref{eq:RPr}.
    \item Based on the requested service reliability ($R_{rel}$), as Eqn. \eqref{eq:Rrel}.
    \item Time taken by the \ac{DDQL} model to find an appropriate node for the placement ($J^{x}_{y}$), Eqn. \eqref{eq:Rplacement} represents the placement reward function ($R_{placement}$).
\end{enumerate}

Algorithm \ref{algo1} describes the overall model. 




\begin{algorithm}[!]\footnotesize
\SetAlgoLined
Initialize DDPG Model: Critic, Actor, Target Critic, Target Actor Networks, \ac{DDPG} Replay Buffer$B_{DDPG}$\\
Initialize \ac{RR} Memory Buffer $B_{RR}$\\
    \ForEach{episode i = 1... epi} 
    {
    Reset the Environment\\
    Initialize substrate node resource $R_H$, substrate link $R_N$ \\
    Received arriving T services \\
        \While{for all T services}
        {
        Using \ac{DyPr} method; online-\ac{RR} model \\
        \eIf{status = "Train"} 
            {\eIf{Transition $<$ Threshold Transition}
                {Priority is selected randomly}
                {Online-$\ac{RR}$ model gets trained}
            }
            {Prediction: using Trained model}
            
        Using \ac{AdSch} method; \ac{DDPG} model\\
            Achieve the Rank for each services \\
        }
    Sort the T service in ascending order (T'), according to achieved rank \\
    \ForEach{time-step t' = 1...T'}
    {
        Admission Control  \\   
        Deployment initiate \\
    }
    } 
\caption{DyPr and AdSch }\label{algo1}
\end{algorithm}
\section{Simulation results}\label{sec:4} 

In this section, the effectiveness of the \ac{DDPG} and \ac{RR} models for \ac{AdSch} and \ac{DyPr} are examined under various conditions, for NetRail (7 nodes, 10 links) and BtEurope (24 nodes, 37 links) topologies, under diverse substrate nodal and link capacities (i.e., from highly available resource topology to easily exhausted). The online services are constructed using the Erdős--Rényi model with different structural complexity and resource requirements. 
Each run consists of 2000 episodes, and each episode is expected to have a maximum of 100 services. 
The \ac{DDPG} and Ridge model is designed in Python language using the PyTorch library, and the simulations are run on an Intel Core i7 processor with 64 GB RAM. Table \ref{table:DDPG} lists the parameters applied to develop the models and services. 
%
\begin{table}[h]
\caption{Parameters}
\begin{tabular*}{\hsize}{@{}@{\extracolsep{\fill}}cc@{}}
 \hline
 \multicolumn{2}{c}{\textbf{DDPG Model}} \\
 \hline
 Alpha & 0.0001 \\
 Beta & 0.001 \\
 Gamma & 0.99 \\
 Batch Size & 64 \\
 Optimizer & Adam \\
 Memory Size & 50000 \\
 Hidden Layers & 6 \\ 
 Neurons per Layer & 300 \\ 
 Neural Network & 2 \\
 Activation function & Sigmoid \\
 \hline
\end{tabular*}
\label{table:DDPG}
\end{table}

In this work, we are considering a `worse-case' scenario, where maximum of 100 services arrive at once, imposing a significant load and high variation on the topology. Moreover, most arriving services need to be deployed sooner, as their threshold waiting time is considerably less, which adds to the system's complexity. Our model's efficiency is compared with traditional queuing models like \ac{FIFO}, \ac{WFQ}, and High-Priority-based scheduling. 
\subsection{Need for Priority}
\begin{figure}
\centering
\begin{center}
    \includegraphics[width=0.7\columnwidth]{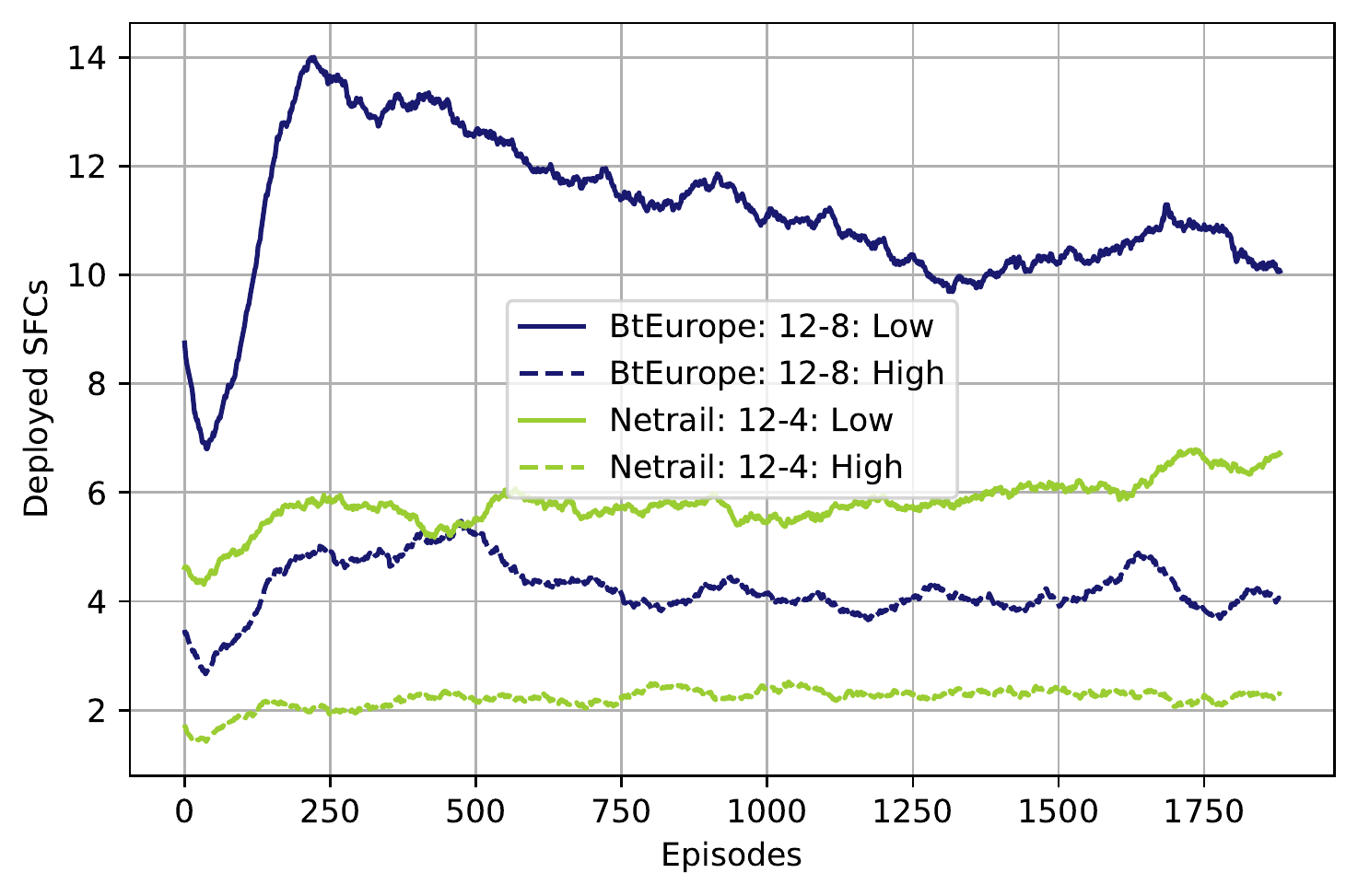}
    \caption{Performance of FIFO}
    \label{fig:FIFO_netrail}
\end{center}
\end{figure}

Figures~\ref{fig:FIFO_netrail} represents the deployment of high and low-priority \acp{SFC} in a \ac{FIFO} manner for Netrail and BtEurope. The model deploys the services as it comes, unable to distinguish between emergency and non-emergency services as the deployment occurrs without any guidelines or prior knowledge of the service. As a result, less than 6\% of urgent/emergency services are preferred, causing considerable rejection of them. This shows the need for priority which is predicted by the online-\ac{RR} model. %
%
%
\begin{figure}[h]
\centering
\begin{center}
 \includegraphics[width=0.7\columnwidth]{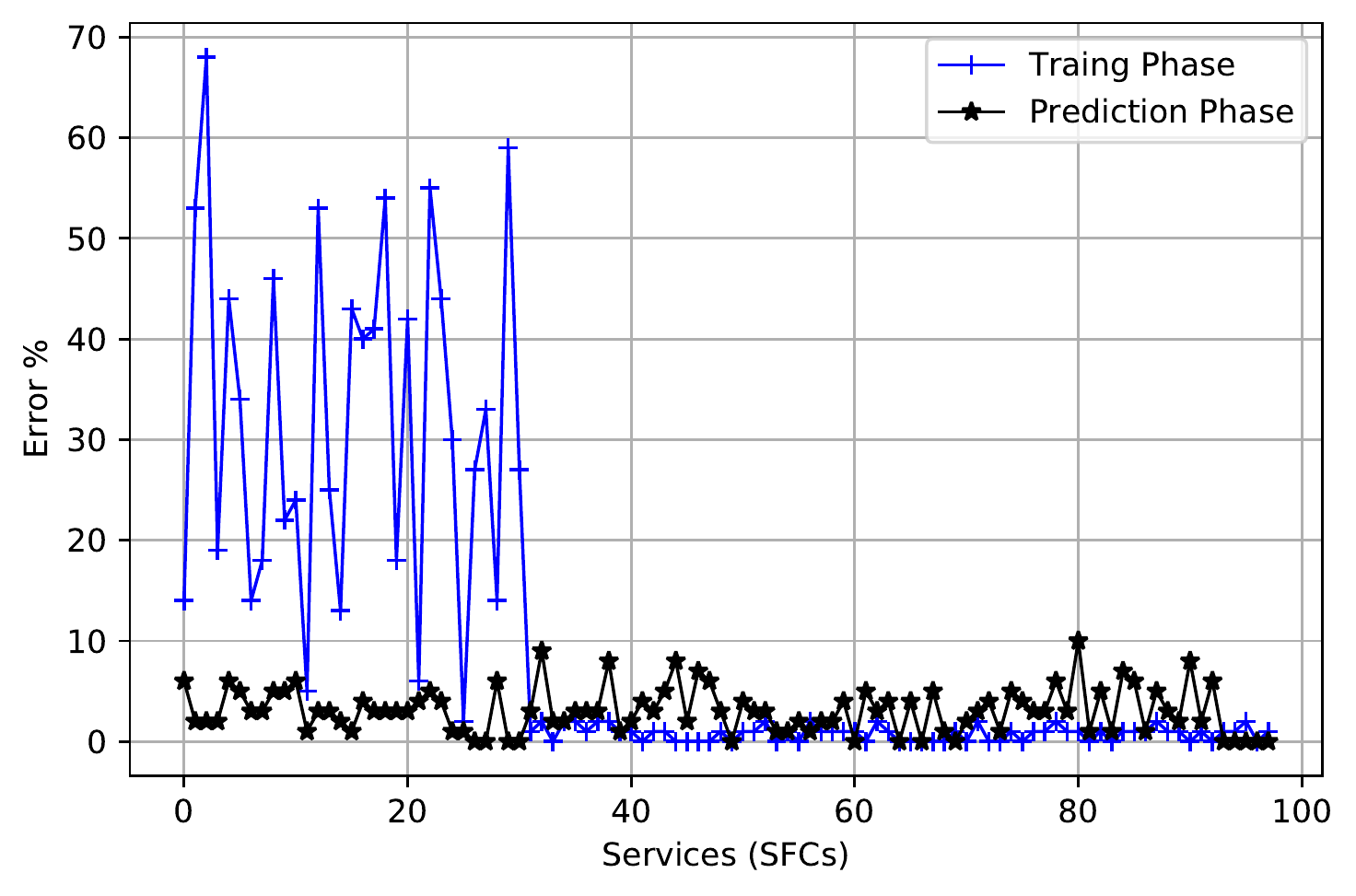}
\caption{Performance of Online Ridge Regression Model}
\label{fig:Training}   
\end{center}
\end{figure}

Figure~\ref{fig:Training} shows the training phase and Prediction phase of the online-\ac{RR} model. The model gets trained until its accuracy exceeds 80\%, however we only displayed for episode 0 for training phase. Initially, the model observed the \ac{SFC} till 32 iterations; later, it commenced learning by discovering a logistical approach. Figure~\ref{fig:Training} depicts the prediction for last episode of a run. It is evident that the model performed well in predicting the priority, as the error\% between the predicted and target priority values are less than 10\%. Thus, the predicted model's accuracy was also above 80\%.

\subsection{Existence of Starvation}
\begin{figure}
\centering
 \includegraphics[width=0.7\columnwidth]{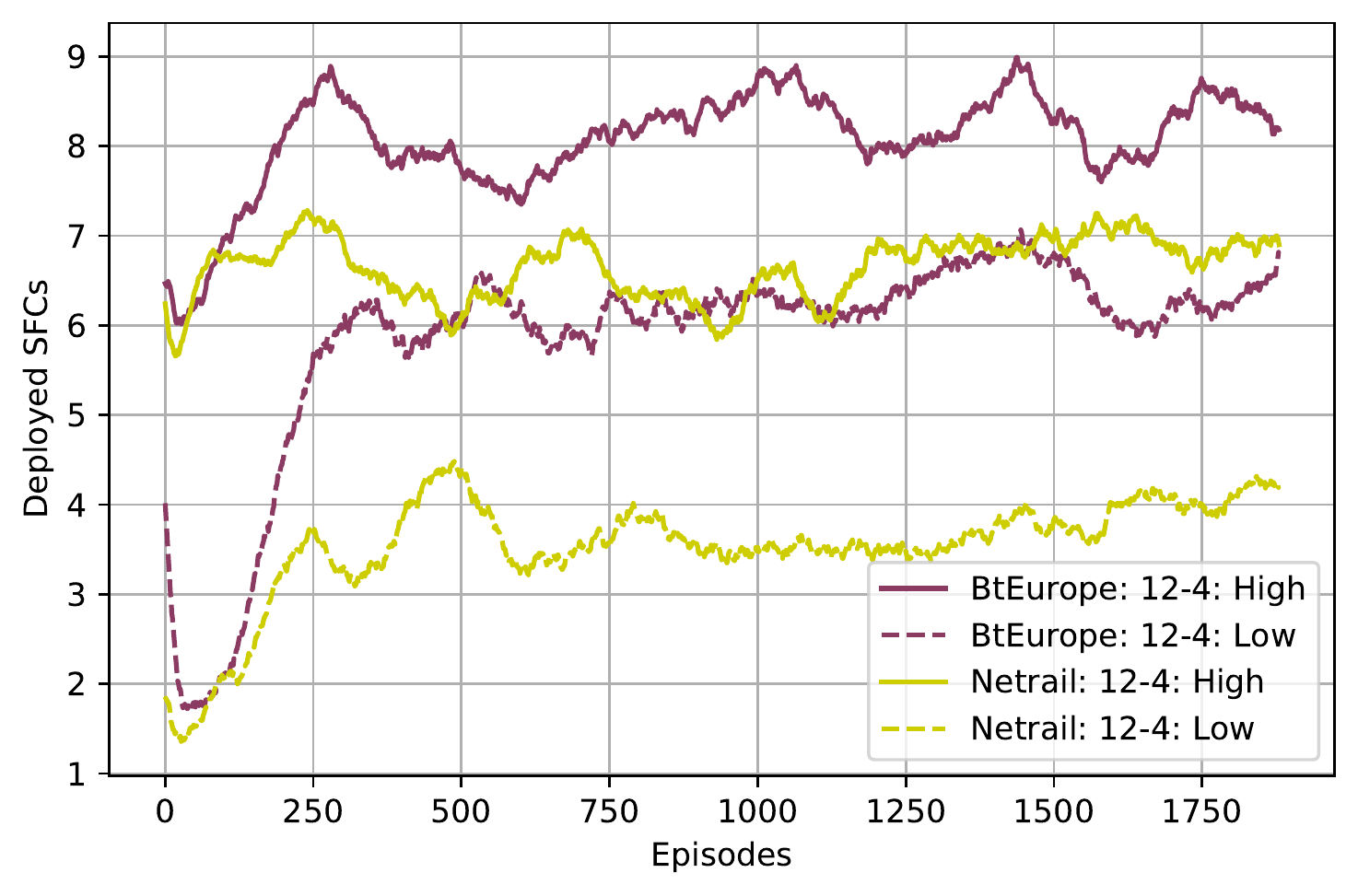}
\caption{Priority Algorithm Netrail SAR: 12-4 scenario}
\label{fig:netrail_12_4_SAR}   
\end{figure}

The effectiveness of the high-priority-based scheduling paradigm is seen in Fig.~\ref{fig:netrail_12_4_SAR}. Here, the model favours high-priority \acp{SFC} above others to avoid the problems caused by \ac{FIFO}. This triggers a prolonged wait for low-priority \acp{SFC} to be deployed, creating a `Starvation' experience and a low acceptance rate. Even with the large nodal resource or higher density topology, as seen in Figure~\ref{fig:netrail_12_4_SAR}, starvation still exists. This starvation gap might slightly get reduced over time for much denser topologies with ample available resources. However, this is unrealistic topology due to the dynamism, where the plentiful resources is not always available.
%
\subsection{Diminish of Starvation}
\begin{figure}[h]
    \centering
    \includegraphics[width=0.7\columnwidth]{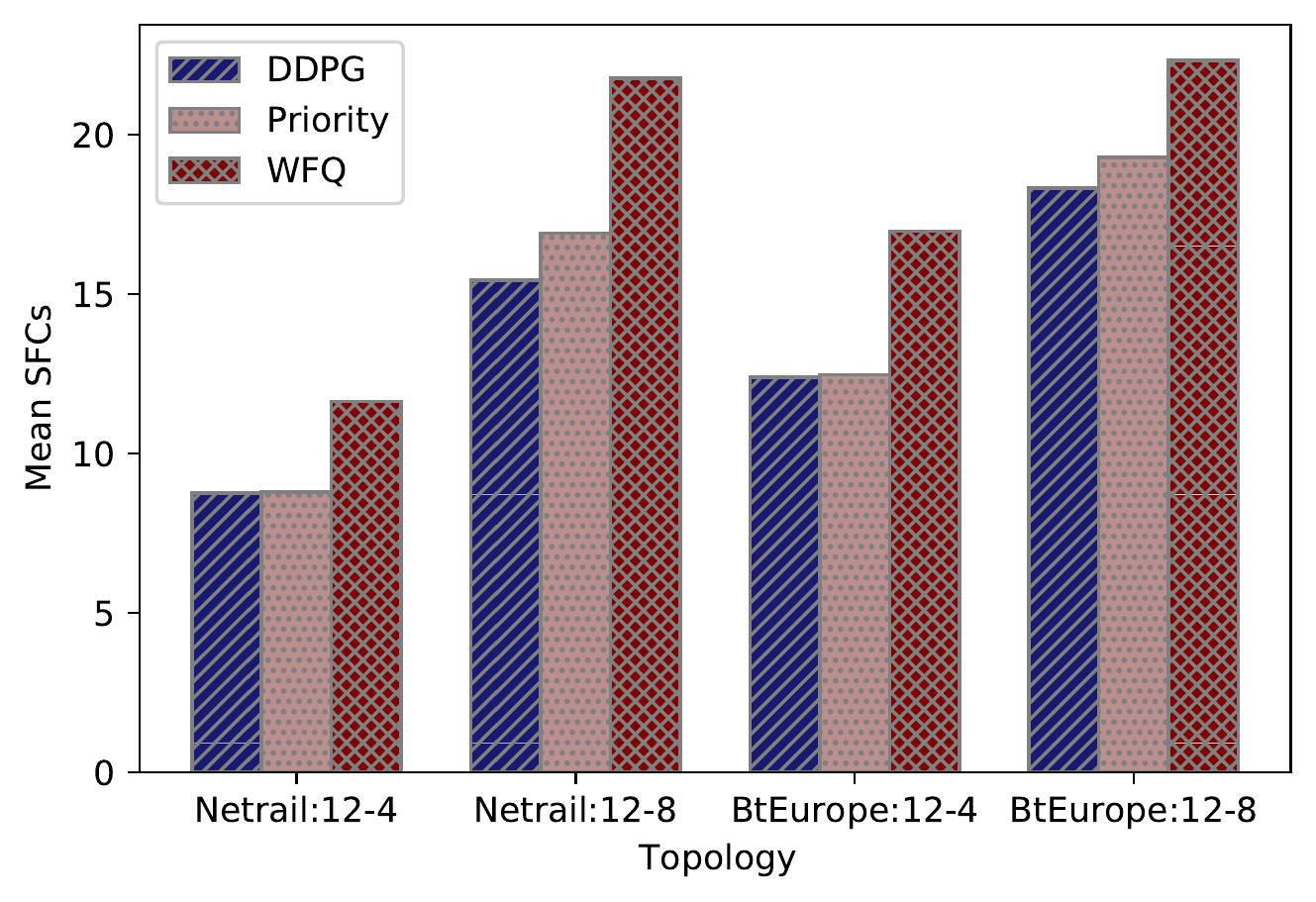}
    \caption{SAR: Deploying Beneficial services}
    \label{fig:DDPG_HM}
\end{figure}

\begin{figure}[h]
    \centering
    \includegraphics[width=0.7\columnwidth]{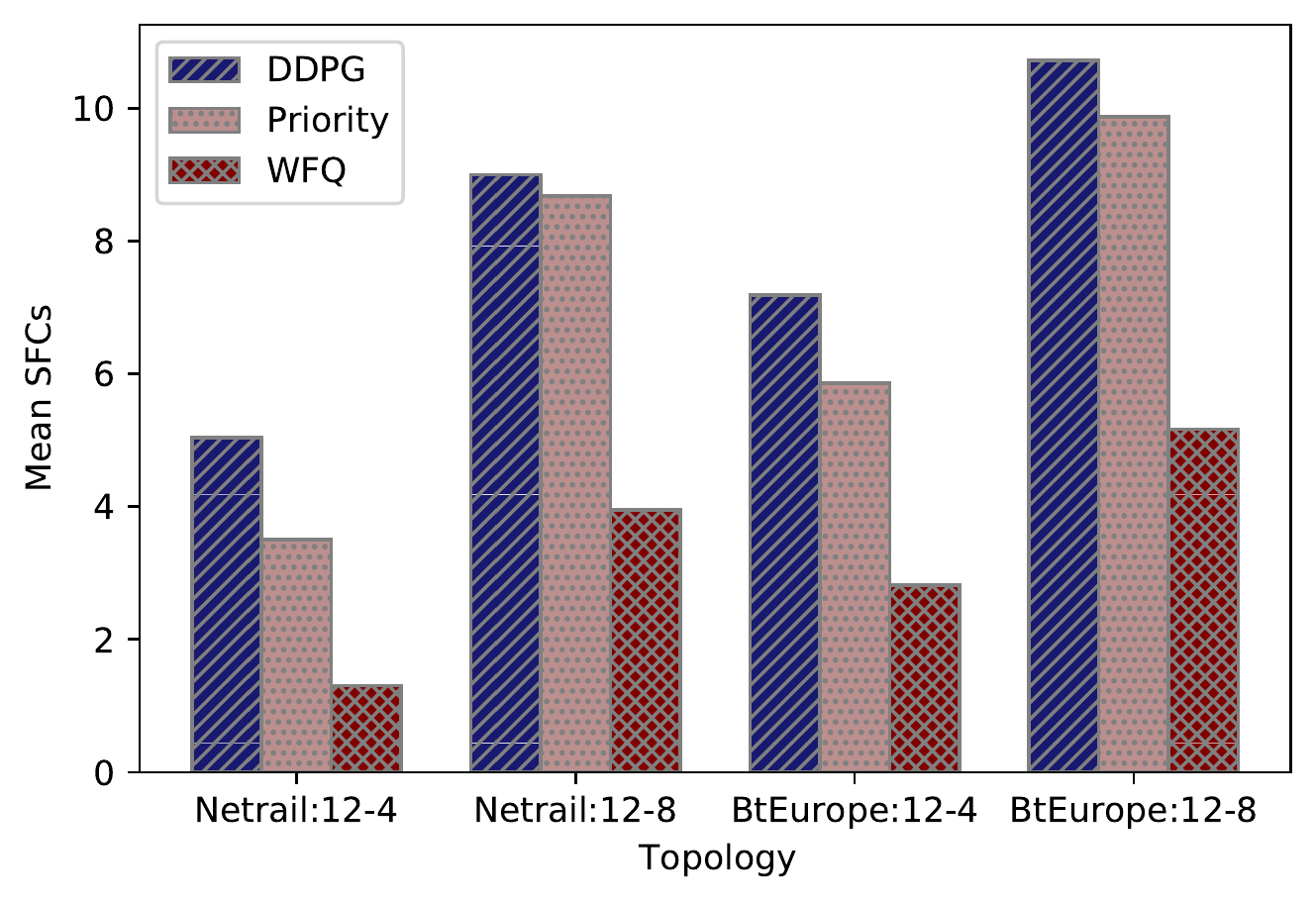}
    \caption{SAR: Deploying Starving services}
    \label{fig:DDPG_Star}
\end{figure}

Figure \ref{fig:DDPG_HM} represents the \ac{SAR} (contains all priority levels, excluding lower-priority \acp{SFC}), and Figure \ref{fig:DDPG_Star} depicts the \ac{SAR} exclusive for low-priority \acp{SFC} for Netrail and BtEurope topologies with 12-4 and 12-8 CPU cores. In Figure \ref{fig:DDPG_HM}, the \ac{WFQ} and high-priority-based models embed a large number of high-priority \acp{SFC} to establish a deploying rule. This pattern is repeated when topological density or nodal capacity increases. From the figures, the \ac{WFQ} and high-priority-based models established a deploying rule by embedding only beneficial \acp{SFC}. This pattern is repeated when topological density or nodal capacity increases. 
The \ac{DDPG} model, on the other hand, discovered a trade-off between high and low-priority \acp{SFC} by identifying `Beneficial and Starving' \acp{SFC}, resulting in a higher rate of deployment of low-priority \acp{SFC} than other models. \ac{DDPG}, like Netrail 12-4 CPU cores, was able to deploy five times more low-priority \acp{SFC} than \ac{WFQ} at the expense of three less high-priority \acp{SFC}. However, in the remaining scenarios (Netrail 12-8; BtEurope 12-4, and BtEurope 12-8), there is a 100\% increase in the deployment of low-priority services (Fig.\ref{fig:DDPG_Star}).
This affected high-priority service deployment, with a 30\% reduction in Netrail and a 25\% reduction in BtEurope (Fig.\ref{fig:DDPG_HM}) respectively. 


\section{Conclusions}\label{sec:5}
The main aim of this study was to showcase the existence of the starvation problem, which the researchers have neglected. In this work, we have explained the existence of the starvation problem in the current scheduling methods and how the scheduling process should not be based only on priority but also on other important factors like threshold waiting time and reliability. With a motive to propose an intelligent scheduling scheme, our model \ac{DDPG} has performed efficiently by deploying twice as many low-priority services as others. The \ac{DDPG} agent successfully identified the `Beneficial and Starving' services, which caused a reduction in the starvation of low-priority services. Moreover, we have proposed a method to define dynamic priority for the upcoming services without hindrance. Our online-\ac{RR} model learns the pattern within 100 transitions, and the accuracy rate for the prediction model is above 80\%. The presence of these problems can have a negative impact in the future if not addressed correctly.
%
\begin{acronym} 
\acro{5G}{the fifth generation of mobile networks}
\acro{6G}{sixth generation of mobile networks}
\acro{ACO}{Ant Colony Optimization}
\acro{AI}{Artificial Intelligence}
\acro{AR}{Augmented Reality}
\acro{ANN}{Artificial Neural Network}
\acro{AdSch}{`Adaptive Scheduling’}
\acro{BB}{Base Band}
\acro{BBU}{Base Band Unit}
\acro{BER}{Bit Error Rate}
\acro{BS}{Base Station}
\acro{BW}{Bandwidth}
\acro{BE}{Best-Effort}
\acro{CC}{Chain Composition}
\acro{C-RAN}{Cloud Radio Access Networks}
\acro{CAPEX}{Capital Expenditure}
\acro{CoMP}{Coordinated Multipoint}
\acro{CR}{Cognitive Radio}
\acro{COC}{Computation Oriented Communications}
\acro{CAeC}{Contextually Agile eMBB Communications}
\acro{D2D}{Device-to-Device}
\acro{DA}{Digital Avatar}
\acro{DAC}{Digital-to-Analog Converter}
\acro{DAS}{Distributed Antenna Systems}
\acro{DBA}{Dynamic Bandwidth Allocation}
\acro{DNN}{Deep Neural Network}
\acro{DC}{Duty Cycle}
\acro{DyPr}{`Dynamic Prioritization'}
\acro{DL}{Deep Learning}
\acro{DSA}{Dynamic Spectrum Access}
\acro{DT}{Digital Twin}
\acro{DRL}{Deep Reinforcement Learning}
\acro{DQL}{Deep Q Learning}
\acro{DDQL}{Double Deep Q Learning}
\acro{DDPG}{Deep Deterministic Policy Gradient}
\acro{$E^2D^2PG$}{Enhanced Exploration Deep Deterministic Policy}
\acro{ER}{Erdős-Rényi}
\acro{EUB}{Expected Upper Bound}
\acro{EDuRLLC}{Event Defined uRLLC}
\acro{eMBB}{enhanced Mobile Broadband}
\acro{FBMC}{Filterbank Multicarrier}
\acro{FEC}{Forward Error Correction}
\acro{FG} {Forwarding Graph}
\acro{FGE}{FG Embedding}
\acro{FIFO}{First-in-First-out}
\acro{FFR}{Fractional Frequency Reuse}
\acro{FSO}{Free Space Optics}
\acro{GA}{Genetic Algorithms}
\acro{GI}{Granularity Index}
\acro{HAP}{High Altitude Platform}
\acro{HL}{Higher Layer}
\acro{HARQ}{Hybrid-Automatic Repeat Request}
\acro{IoE}{Internet of Everything}
\acro{IoT}{Internet of Things}
\acro{ILP}{Integer Linear Program}
\acro{KPI}{Key Performance Indicator}
\acro{LAN}{Local Area Network}
\acro{LAP}{Low Altitude Platform}
\acro{LL}{Lower Layer}
\acro{LOS}{Line of Sight}
\acro{LTE}{Long Term Evolution}
\acro{LTE-A}{Long Term Evolution Advanced}
\acro{MAC}{Medium Access Control}
\acro{MAP}{Medium Altitude Platform}
\acro{MIMO}{Multiple Input Multiple Output}
\acro{ML}{Machine Learning}
\acro{MME}{Mobility Management Entity}
\acro{mmWave}{millimeter Wave}
\acro{MNO}{Mobile Network Operator}
\acro{MR}{Mixed Reality}
\acro{MILP}{ Mixed-Integer Linear Program}
\acro{MDP}{Markov Decision Process}
\acro{mMTC}{massive Machine Type Communications}
\acro{NAI}{Network Availability Index}
\acro{NASA}{National Aeronautics and Space Administration}
\acro{NAT}{Network Address Translation}
\acro{NN}{Neural Network}
\acro{NF}{Network Function}
\acro{NFP}{Network Flying Platform}
\acro{NTN}{Non-terrestrial networks}
\acro{NFV}{Network Function Virtualization}
\acro{NS}{Network Service}
\acro{OFDM}{Orthogonal Frequency Division Multiplexing}
\acro{OSA}{Opportunistic Spectrum Access}
\acro{PAM}{Pulse Amplitude Modulation}
\acro{PAPR}{Peak-to-Average Power Ratio}
\acro{PGW}{Packet Gateway}
\acro{PHY}{physical layer}
\acro{PSO}{Particle Swarm Optimization}
\acro{PT}{Physical Twin}
\acro{PU}{Primary User}
\acro{Pr}{Premium}
\acro{QAM}{Quadrature Amplitude Modulation}
\acro{QoE}{Quality of Experience}
\acro{QoS}{Quality of Service}
\acro{QPSK}{Quadrature Phase Shift Keying}
\acro{QL}{Q-Learning}
\acro{RA}{Resource Allocation}
\acro{RF}{Radio Frequency}
\acro{RN}{Remote Node}
\acro{RRH}{Remote Radio Head}
\acro{RRC}{Radio Resource Control}
\acro{RRU}{Remote Radio Unit}
\acro{RL}{Reinforcement Learning}
\acro{RR}{Ridge Regression}
\acro{SCH}{Scheduling}
\acro{SU}{Secondary User}
\acro{SCBS}{Small Cell Base Station}
\acro{SDN}{Software Defined Network}
\acro{SFC}{Service Function Chaining}
\acro{SLA}{Service Level Agreement}
\acro{SNR}{Signal-to-Noise Ratio}
\acro{SON}{Self-organising Network}
\acro{SAR}{Service Acceptance Rate}
\acro{TDD}{Time Division Duplex}
\acro{TD-LTE}{Time Division LTE}
\acro{TDM}{Time Division Multiplexing}
\acro{TDMA}{Time Division Multiple Access}
\acro{TWT}{Threshold Waiting Time}
\acro{UE}{User Equipment}
\acro{UAV}{Unmanned Aerial Vehicle}
\acro{URLLC}{Ultra-Reliable Low Latency Communications}
\acro{USRP}{Universal Software Radio Platform}
\acro{VL}{Virtual Link}
\acro{VNF}{Virtual Network Function}
\acro{VNF-FG}{VNF-Forwarding Graph}
\acro{VNF-FGE}{VNF-FG Embedding}
\acro{VR}{Virtual Reality}
\acro{WFQ}{Weighted Fair Queuing}
\acro{XAI}{Explainable Artificial Intelligence}
\end{acronym}
\bibliographystyle{IEEEtran}
\bibliography{myref.bib}
\end{document}